\documentclass[a4paper,12pt]{article}
\usepackage{graphicx}
\usepackage{wrapfig}
\usepackage{amsmath}

\newcommand{\mhf}{m_{1/2}}
\newcommand{\ma}{m_A}
\newcommand{\mzr}{m_0}
\newcommand{\Azr}{A_0}

\newcommand{\n}{\tilde{\chi}_1^0}
\newcommand{\simless}{\hspace{0.3em}\raisebox{0.4ex}{$<$}\hspace{-0.75em}\raisebox{-.7ex}{$\sim$}\hspace{0.3em}}
\renewcommand{\partial}{\del}

\begin{document}
\begin{titlepage}
\begin{center}
{\large \bf
Neutralino Dark Matter in minimal
supersymmetric standard model with natural light Higgs sector}
\vspace{6ex}
\\

\renewcommand{\thefootnote}{\alph{footnote}}
S.-G.~Kim\footnote{e-mail: sunggi@eken.phys.nagoya-u.ac.jp}, 
N.~Maekawa\footnote{e-mail: maekawa@eken.phys.nagoya-u.ac.jp},
K.~I.~Nagao\footnote{e-mail: nagao@eken.phys.nagoya-u.ac.jp},\\ 
K.~Sakurai\footnote{e-mail: sakurai@eken.phys.nagoya-u.ac.jp}, 
T.~Yoshikawa\footnote{e-mail: tadashi@eken.phys.nagoya-u.ac.jp}

\vspace{4ex}
{\it Department of Physics, Nagoya University, Nagoya 464-8602, Japan}\\

\end{center}

\renewcommand{\thefootnote}{\arabic{footnote}}
\setcounter{footnote}{0}
\vspace{6ex}

\begin{abstract}
We study the neutralino relic density in the minimal supersymmetric standard model
with natural light Higgs sector in which all Higgs masses, the supersymmetry
(SUSY) breaking 
parameters, and the higgsino mass parameter $\mu$ are of order the weak scale.
To realize this situation we adopt nonuniversal Higgs masses at the grand unified
scale.
We show that in some parameter space in which the SUSY breaking parameters are 
comparatively small, not only the constraint from the observed
relic density of dark matter but also the LEP Higgs bound and  
the constraint from the $b\rightarrow s\gamma$ process are satisfied. 
In most of the parameter space, the neutralino relic density becomes smaller 
than 
the observed relic density
in contrast with the results in the constrained minimal 
SUSY standard model (CMSSM).
The reason is that the neutralino coannihilation processes to Higgs bosons open 
even if the gaugino mass is small and 
the cross sections become large due to the small dimensionful parameters.
Especially small $\mu$ parameter and the light CP-odd Higgs, which are difficult 
to be realized in the CMSSM,
are essential for the result.

\end{abstract}
\end{titlepage}

\section{Introduction}
The minimal supersymmetric (SUSY) standard model (MSSM) is one of the hopeful extensions of the standard model (SM).
It is attractive not only in the point of the weak scale stability, 
but also in the fact that
SUSY models with R-parity have the lightest supersymmetric particle (LSP)
as a good candidate for dark matter.
Since the thermal relic density of the LSP can be calculated once we
fix the parameters in the MSSM, it is interesting to examine 
parameter space, which
is consistent with the observed value 
\begin{eqnarray}
\Omega_{\mathrm{DM}} {\rm h}^2&=&\Omega_{m} {\rm h}^2-\Omega_{b} {\rm h}^2\\
&=&\left(0.1277^{+0.0080}_{-0.0079}\right)-\left(0.02229\pm 0.00073\right),
\label{omega_dm}
\end{eqnarray}
where $\Omega_{\mathrm{DM}}, \Omega_m$ and $\Omega_b$ are
the energy densities of
dark matter,
matter, 
and baryon of the universe \cite{Spergel:2006hy}.
Here, ${\rm h}$ is the normalized Hubble parameter such as 
the present Hubble constant is given by
$H_0=100{\rm h}\,\mathrm{km}\mathrm{s^{-1}}\mathrm{Mpc^{-1}}$ .
Many studies have been done about the neutralino relic density in the 
constrained MSSM (CMSSM) \cite{Ellis:2001msa}, 
in which all dimensionful parameters can be
presented by only five parameters, the unified gaugino mass $m_{1/2}$,
the unified scalar mass $m_0$, the universal couplings for three scalar vertex $A$, the parameter for Higgs mixing $B$, and the higgsino mass
$\mu$.
Unfortunately the allowed region for the parameters in the CMSSM are 
quite limited 
because in most of the parameter region consistent with experiments,
 the calculated 
thermal relic density of the neutralino become too large to satisfy the
observed value.
Moreover, in the CMSSM, 
the LEP constraint to the standard model 
Higgs mass, $m_h>114.4$\,GeV ($95\%$ C.L.) \cite{Higgsmassbound},
requires comparatively large SUSY 
breaking parameters in order to make the lightest MSSM Higgs heavier
via loop corrections than the upper bound for the SM Higgs mass
$m_h>114.4$\,GeV.
Such large SUSY breaking parameters destabilize the weak scale. 
This problem is called as the little hierarchy problem. 

Recently, it has
been pointed out that in the nonuniversal Higgs mass model, the LEP 
constraints can be avoided due to the smaller $ZZh$ coupling than in 
the SM \cite{Kane:2004tk, Drees:2005jg, kim, Belyaev:2006rf}.
Here,
$h$ is the lightest CP even Higgs.
Therefore, the large SUSY breaking parameters are not needed. 
This avoids the little hierarchy problem. To obtain the small $ZZh$
coupling, generically, the light 
Higgs sector is required, in which not only the usual CP even Higgs 
but also the other Higgs bosons  have the weak scale 
masses. 
Contributions of the light charged Higgs to the $b\rightarrow s\gamma$
 process would be too large to be consistent with the experimental value,
if the chargino contribution has not compensated the charged 
Higgs contribution. 
This cancellation due to the supersymmetry works well \cite{Ferrara:1974wb, Bertolini:1990if, Oshimo:1992zd, kim}
 because all the mass scales in the Feynman diagrams contributing
 the $b\rightarrow s\gamma$ process are of order
 the weak scale in the models with the natural SUSY breaking parameters
and the light Higgs sector.
In such models all the dimensionful parameters are of order the weak scale.
The charged Higgs mass $m_{H^\pm}$ and
the Higgsino mass $\mu$ are fixed at the weak scale.
The sfermion mass $\mzr$,
the gaugino mass $\mhf$
and the scalar three point coupling $\Azr$
are fixed at the GUT scale.
We call such scenario the natural light Higgs scenario.

In this paper, we calculate the thermal relic density of the lightest
neutralino in the natural light Higgs scenario. 
The result is totally different from the result in the CMSSM.
The thermal relic density in this scenario tends to be
smaller than the observed dark matter energy density.
This is mainly because 
the neutralino coannihilation modes to Higgs bosons
such as $\n\n\to hA$ and $\n\n \to HA$
open due to the light Higgs sector and because
the cross sections become large due to the
small dimensionful parameters, especially small $\mu$.
Here $\n$, $A$, and $H$ are the lightest neutralino,
the CP odd Higgs, and the heaviest CP even Higgs,
respectively.
The  larger total annihilation cross section of the neutralino
leads to the smaller thermal relic density.
If $\mu$ is large,
the cross sections of neutralino coannihilation processes to 
Higgs bosons decrease
because of the small Higgsino components of the lightest neutralino.
Thus the energy density of the neutralino becomes larger than 
that with small $\mu$. 

There are two essential points.
One of them is that the light Higgs bosons make it possible to open the neutralino coannihilation processes to Higgs bosons.
And the other is that the small $\mu$ parameter makes the cross sections large. 
In the CMSSM, it is
difficult to satisfy both of them. Roughly speaking, to obtain the light
Higgs sector, the both mass parameters $m_1^2=m_{H_d}^2+|\mu|^2$ and 
$m_2^2=m_{H_u}^2+|\mu|^2$ in the Higgs potential
\begin{equation}
V=m_1^2|H_d|^2+m_2^2|H_u|^2+(m_3^2H_dH_u+h.c.)+\frac{g^{\prime 2}+g
^2}{8}
(|H_d|^2-|H_u|^2)^2
\end{equation}
must be around the weak scale, which is difficult to be satisfied in
the CMSSM because $m_1^2-m_2^2$ becomes much larger than the weak scale.
Here, $H_u$ and $H_d$ are up-type Higgs and 
down-type Higgs, respectively. 
Actually, in the CMSSM, the difference
$m_1^2-m_2^2$ at the weak scale becomes roughly 3$\mhf^2$,
where $\mhf$ is taken roughly larger than 300\,GeV to satisfy the LEP constraint
to the SM Higgs mass bound.

For the reasons stated above,
the neutralino relic density in the natural light Higgs scenario tends to be 
smaller than the observed energy density.
And,
in this scenario,
there are parameter regions where the neutralino relic density 
agrees with the observation.

In section \ref{sec:neumerical_analyses}, we show the numerical 
calculation of the
neutralino thermal relic density in the natural light Higgs scenario.
After a discussion about the allowed region,
we conclude in section \ref{sec:summary}.

\section{Neutralino relic density in natural light Higgs scenario
\label{sec:neumerical_analyses}}
In this section, we calculate the neutralino thermal relic density 
numerically in the MSSM with the light Higgs sector and natural SUSY
breaking parameters. 
There are two additional dimensionful parameters,
 $m_{H_u}^2$ and $m_{H_d}^2$ in the nonuniversal Higgs mass model. 
Then we have seven parameters. One of seven parameters is fixed by the
Z boson mass, and thus we have six parameters.
In this paper,
three of them, the universal sfermion mass $\mzr$,
the gaugino mass $\mhf$, 
and the universal coupling for the three scalar interaction $\Azr$
are fixed at the GUT scale, and the other parameters,  $\mu$ parameter,
the ratio of two Higgs vacuum expectation values
$\tan{\beta}$, and the CP odd Higgs mass $m_A$ are fixed at the weak 
scale.

In this paper, 
we adopt the small $ZZh$ coupling scenario
\cite{Kane:2004tk, Drees:2005jg, kim, Belyaev:2006rf}
to satisfy the LEP constraints to the SM Higgs mass\footnote{
If we adopt the nonuniversal sfermion masses, then the naturalness
requires that only the masses of the stops must be around the weak scale
and the other sfermion masses can be taken much larger than the weak
scale.
$E_6$ GUT with horizontal symmetry naturally obtains such nonuniversal
sfermion masses, and our discussion in this paper can be applied to
the nonuniversal sfermion mass model
\cite{Maekawa:2002eh}.}.
We fix some of the parameters, $\Azr$, $\mu$, $\tan\beta$, and $m_A$
 to reduce the number of parameters.
$m_A$ and $\tan\beta$ are important to realize small ZZh coupling, because
CP-even Higgs mass matrix is roughly given by
\begin{eqnarray}
\left(
\begin{array}{cc}
m_A^2 & -(m_A^2+m_Z^2)\sin\beta\cos\beta \cr
            -(m_A^2+m_Z^2)\sin\beta\cos\beta & m_Z^2+\delta m_{H_u}^2
            \end{array}
            \right)
\end{eqnarray}
when $\tan\beta\gg 1$. 
Here, $\delta m_{H_u}^2$ is the loop correction to
$m_{H_u}^2$ which can be large due to the large top Yukawa coupling.
When $m_A^2<m_Z^2+\delta m_{H_u}^2$, the main component of the lightest Higgs
$h$ becomes $H_d$ which has only very small $ZZH_d$ coupling. Moreover,
when $m_Z^2+\delta m_{H_u}^2-m_A^2\gg(m_A^2+m_Z^2)\sin\beta\cos\beta$,
 $h$ includes only small component of $H_u$. 
On the other hand,
if the off-diagonal element is large (i.e., $\tan\beta$ is small) and/or the
difference of the diagonal element is small (i.e. $m_A$ is large or the loop
correction is small), the $H_u$
component in $h$ becomes large, which results in large $ZZh$ coupling.
Therefore,
as discussed in \cite{kim}, $7<\tan\beta$ and $90$\,GeV$<m_A<110$\,GeV are required (If CP-even Higgs mass is smaller than 
90\,GeV, the $Z\rightarrow Ah$ process gives a severe constraint by
LEP experiments.).
In this paper, we
take rather large $\tan\beta$ and small $m_A$ as $\tan\beta=15$ and  $m_A=96$
\,GeV which satisfy that the ZZh coupling is
smaller than a half of the SM ZZh coupling in the parameter region discussed in
this paper.\footnote{We checked by \textit{FeynHiggs2.6.4} \cite{feynhiggs} that the $ZZh$ coupling is smaller than half of the SM coupling in the parameter region we took in this paper.}
We examine two cases for $A_0$, $A_0=0$ and $250$\,GeV.
For the $\mu$ parameter, 
we have a strong reason to take it as the weak
scale in the scenario with the small $ZZh$ coupling.
Since the both mass parameters $m_1^2=m_{H_d}^2+|\mu|^2$ and 
$m_2^2=m_{H_u}^2+|\mu|^2$ in the Higgs potential
must be around the weak scale to obtain the small $ZZh$ coupling,
not only the tuning for $m_1^2$ but also that for $m_2^2$
are required if the $\mu$ parameter is much 
larger than the weak scale. 
Therefore, we take $\mu=275$, $300$, $325$, $350$\,GeV,
and for comparison, we examine 
the case $\mu=600$\,GeV, which is not so natural\footnote{We calculated for $\mu=250$\,GeV  and found that the neutralino relic density is 
smaller than the observed value in all natural parameter space.}.

To obtain the low-energy parameters (sfermion masses, A term and gaugino masses) 
from the GUT scale parameters, 
we use the 1-loop renormalization group equations (RGEs) from the GUT 
scale down to the electroweak scale.
In the calculation,
we choose the GUT scale Higgs masses in order to realize  
$m_A=96$\,GeV, $\tan\beta=15$, $\mu=275$, $300$, $325$, $350$\,GeV and $m_Z=91.18$\,GeV
 at the electroweak scale.
Concretely speaking, we choose the GUT scale parameters by calculating 
the RGEs iteratively to realize the parameters we fixed at the low 
energy.
Then we calculate the neutralino relic density
by using \textit{micrOMEGAs1.3.7} \cite{micromegas, micromegas1.3} package
by inputting the parameters at the electroweak scale.
Here we took the on-shell top mass $M_t=172.6$\,GeV \cite{Group:2008nq}.

\begin{figure}[!ht]
\begin{tabular}{cc}
\begin{minipage}{.5\textwidth}
\includegraphics[keepaspectratio,width=0.90\textwidth]{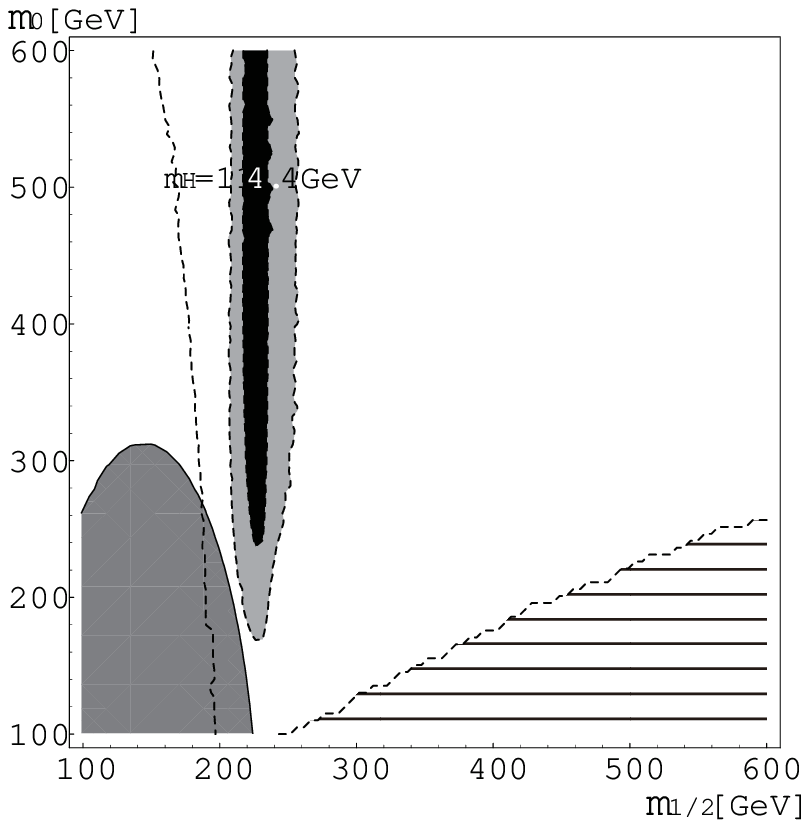}
\caption{$A_0=0$\,GeV, $\mu=275$\,GeV}
\label{fig:a0mu275tb15}
\end{minipage}&
\begin{minipage}{.5\textwidth}
\vspace*{5mm}
\includegraphics[keepaspectratio,width=.88\textwidth]{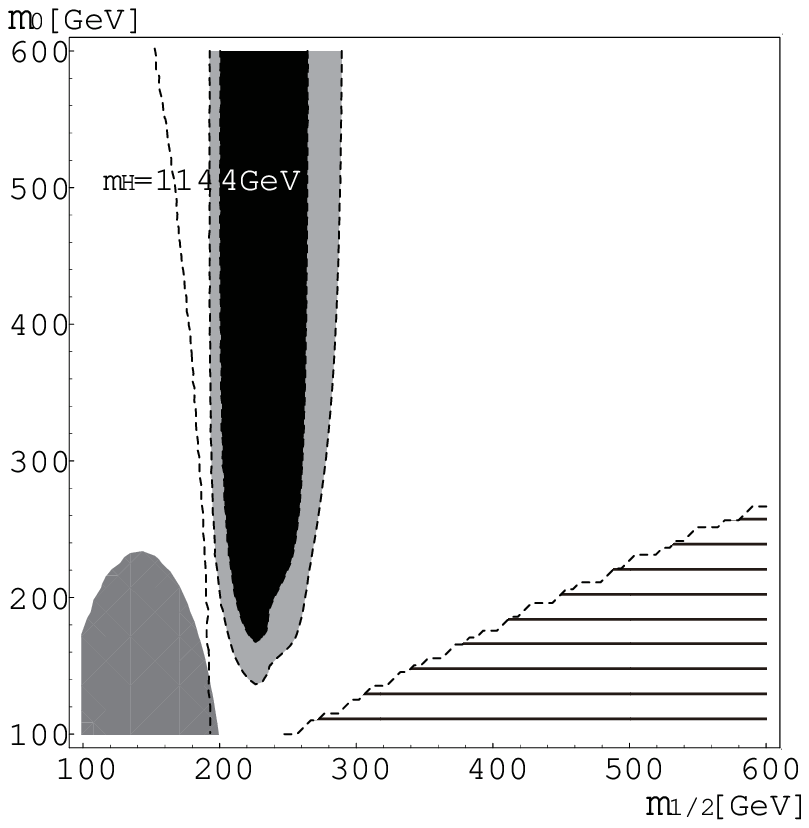}
\caption{$A_0=0$\,GeV, $\mu=300$\,GeV}
\vspace*{5mm}
\label{fig:a0mu300tb15}
\end{minipage}\\
\begin{minipage}{.5\textwidth}
\includegraphics[keepaspectratio,width=0.88\textwidth]{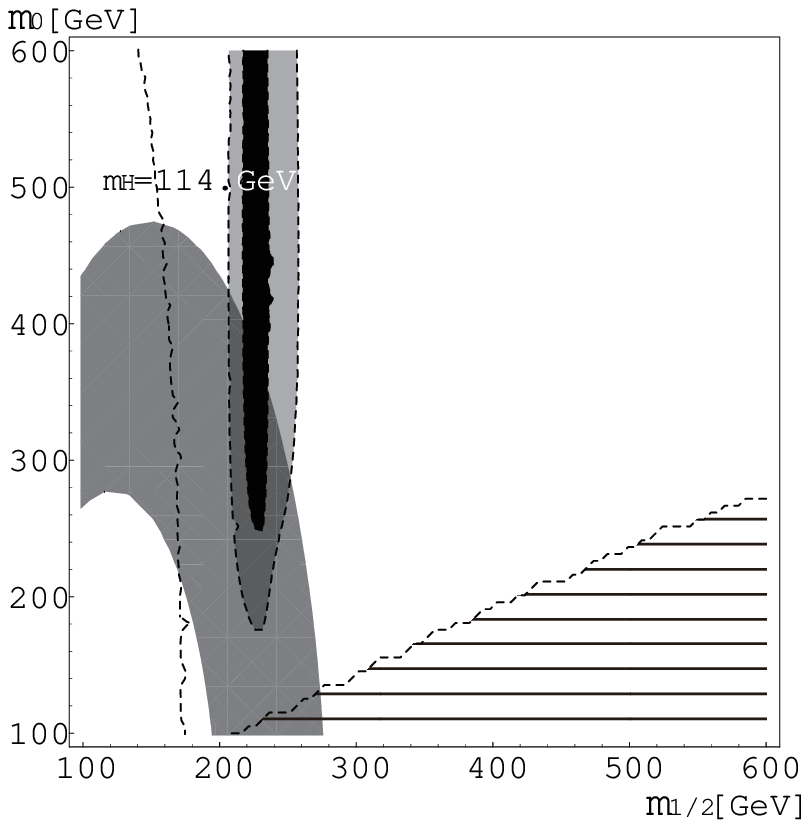}
\caption{$A_0=250$\,GeV, $\mu=275$\,GeV}
\label{fig:a250mu275tb15}
\end{minipage}&
\begin{minipage}{.5\textwidth}
\includegraphics[keepaspectratio,width=.88\textwidth]{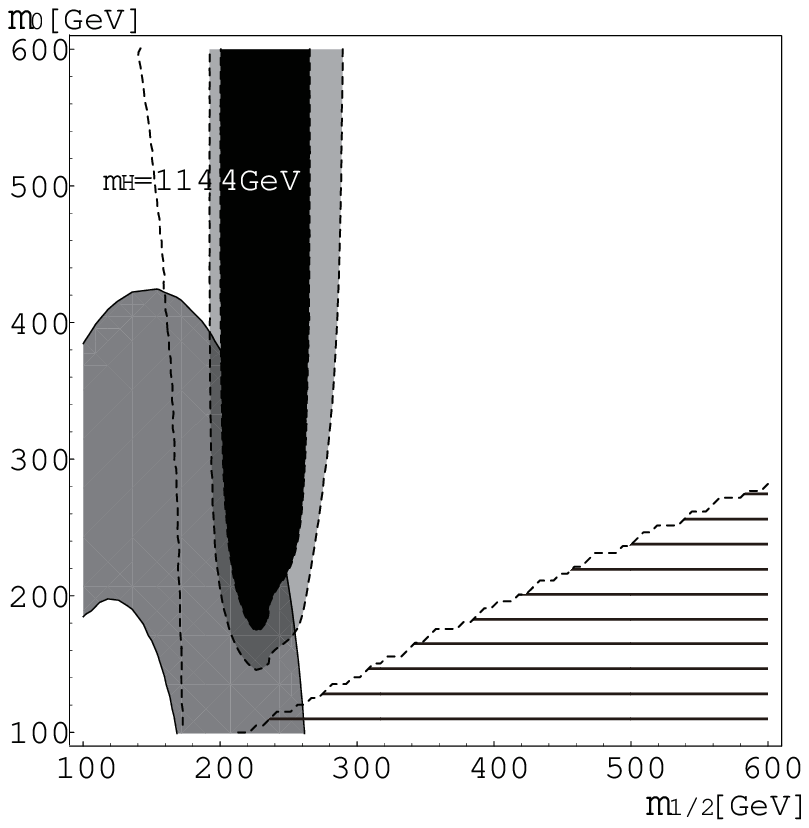}
\caption{$A_0=250$\,GeV, $\mu=300$\,GeV}
\label{fig:a250mu300tb15}
\end{minipage}
\\ \multicolumn{2}{c}{\begin{minipage}{\textwidth}
\vspace*{5mm}
\vbox{
The ($\mhf$, $\mzr$) planes for $\Azr=0$\,GeV (upper), $\Azr=250$\,GeV (lower), $\mu=275$\,GeV (left), $\mu=300$\,GeV (right) with $\tan{\beta}=15$ and $\ma=96$\,GeV. Each light gray area is the region where the relic density is consistent with the current observation. The relic density is larger (smaller) than that of the light gray area in each black (white) region. In the area with horizontal stripes, stau is the LSP. Each dark gray area is the region where $Br(b\rightarrow s\gamma)$ is consistent with the experiment in the sense described in the text. Dashed line is the contour on which the mass of the heaviest CP even Higgs is $114.4$\,GeV.}
\end{minipage}}
\end{tabular}
\end{figure}

We display in Fig.\ref{fig:a0mu275tb15}-Fig.\ref{fig:a250mu300tb15}
the relic density of 
the lightest neutralino in the natural light Higgs scenario. 
In the figures,
the light gray area is  the cosmologically preferred region
where the neutralino relic density is consistent with eq.(\ref{omega_dm}).
The regions with larger and smaller relic density are painted black and white,
respectively.
The horizontal-striped region is excluded because the LSP becomes stau, 
which is a charged particle.
The dark gray areas are allowed regions for $b\to s\gamma$ constraint.
Since the observed branching ratio for the process,
$Br(b\to s\gamma)_{exp.}=(355 \pm26)\times 10^{-6}$ \cite{Barberio:2007cr},
is now in agreement with the SM estimations,
$Br(b\to s\gamma)_{SM}=(315 \pm23)\times 10^{-6}$ \cite{Misiak:2006zs},
$Br(b\to s\gamma)_{SM}=(357 \pm49)\times 10^{-6}$ \cite{Andersen:2006hr}, 
and
$Br(b\to s\gamma)_{SM}=(298 \pm26)\times 10^{-6}$ \cite{Becher:2006pu},
we seek the region where the MSSM contributions for the process are 
moderate.
Here we assume the Minimal Flavor Violation (MFV) and the 
primary contributions coming from the SM, charged Higgs and chargino are 
taken into account using input parameters and RG method described in the 
previous paragraph.
For simplicity, we require the effective Wilson coefficient $C_7$ at b 
quark mass scale to be within 20 percents difference 
from the SM
 prediction in the leading order approximation for the process.
That is to say,
the coefficients evaluated at the 
electroweak scale from 1-loop diagrams are translated into that of b quark scale ($\mu_b=4.7$\,GeV) 
values using $8 \times 8$ evolution matrix 
calculated at 2-loop 
level \cite{Bertolini:1990if, Ciuchini:1993ks}.
Dashed line denotes the heavy Higgs mass bound,
$m_H=114.4$\,GeV. In the model with small $ZZh$ coupling, the $ZZH$
coupling becomes almost the same as the SM value, and therefore,
the LEP constraints to the SM Higgs mass can be roughly applied to
the heaviest Higgs mass in the MSSM.
In all parameter region in these figures, 
the lightest CP-even Higgs mass is 
larger than 90\,GeV.
If $\tan{\beta}$ is fairly large,
the constraint from $B_s \to \mu^+\mu^-$ process must be
taken into account \cite{Babu:1999hn, Isidori:2006pk, Ellis:2007fu}.
However,
the constraint can be negligible 
in the parameter region we took in our calculation,
when $\tan{\beta}\simeq 15$.

\begin{figure}[ht]
\begin{tabular}{cc}
\begin{minipage}{.5\textwidth}
\hspace{-1mm}
\includegraphics[keepaspectratio,width=.88\textwidth]{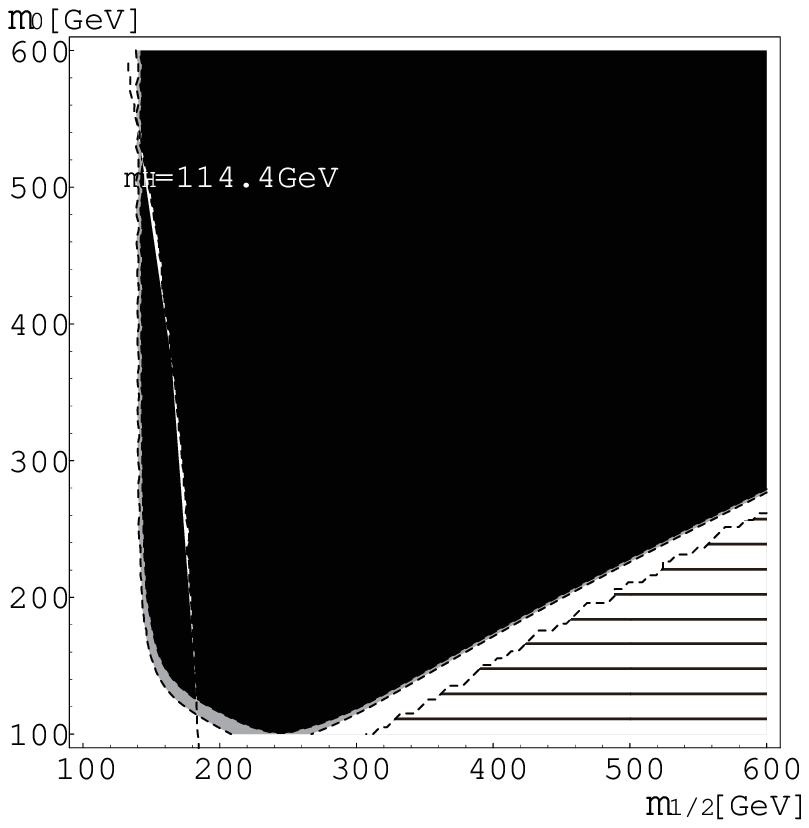}
\caption{$A_0=0$\,GeV, $\mu=600$\,GeV}
\label{fig:a0mu600tb15}
\end{minipage}&
\begin{minipage}{.5\textwidth}
\hspace{-1mm}
\includegraphics[keepaspectratio,width=.88\textwidth]{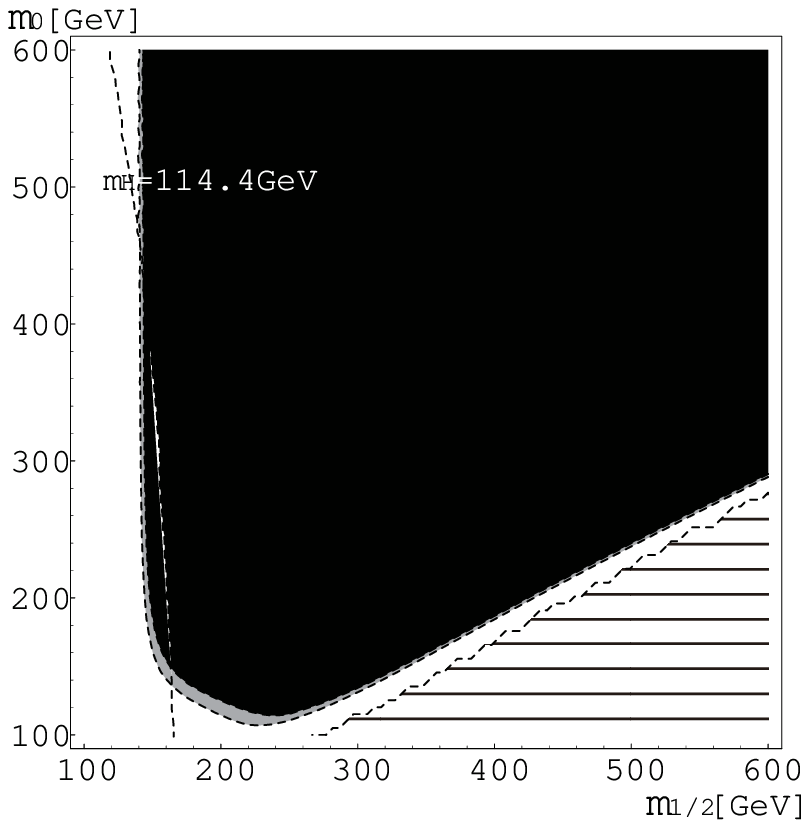}
\caption{$A_0=250$\,GeV, $\mu=600$\,GeV}
\label{fig:a250mu600tb15}
\end{minipage}
\\ \multicolumn{2}{c}{\begin{minipage}{\textwidth}
\vspace{5mm}
\vbox{The ($\mhf$, $\mzr$) planes for $\Azr=0$\,GeV (left) and $\Azr=250$\,GeV (right) with $\mu=600$\,GeV, $\tan{\beta}=15$ and $\ma=96$\,GeV. The usage of each color and line is the same as the previous figures. In these figures, the constraints from $b\rightarrow s\gamma$ are not presented.}
\end{minipage}}
\end{tabular}
\end{figure}

There are two reasons for the small relic density.
For roughly $\mhf \simless 200$\,GeV,
the cross section of processes $\n\n \rightarrow A \rightarrow b\bar{b}$ and $\n\n \rightarrow Z \rightarrow b\bar{b}$ become so large
that the neutralino relic density is smaller than the observed relic density of dark matter.
The neutralino is roughly half as heavy as the CP-odd Higgs and the Z boson,
so the sum of the masses of two neutralinos is nearly on the pole 
of the CP-odd Higgs and the Z boson in this region. 
The cross section of this process decreases 
 as the gaugino mass 
grows up, because the sum of the two neutralinos masses becomes away 
from the poles.
In the CMSSM,
this left preferred areas are mostly excluded by the LEP bound.
On the contrary,
all preferred regions are allowed by the Higgs mass bound in the models
with the small $ZZh$ coupling.  
There is another area in which the neutralino relic density becomes 
smaller than the observed value when the gaugino mass becomes larger.
This is because the modes of $\n\n\to$ two bosons
such as $\n\n \to h A$,
$\n\n \to H A$, 
 and $\n\n \to W^{\pm}H^{\mp}$
 modes open as well as $\n\n \to Z h$.
It is essential that in the natural light Higgs scenario,
all Higgs bosons are light.
In the CMSSM,
the relic density of the region corresponding to this region 
is larger than cosmologically preferred range \cite{ellis}.
We can plot similar graphs even in the models with small $ZZh$ coupling
in the case of large $\mu$ parameter as we show in 
Fig.\ref{fig:a0mu600tb15} and Fig.\ref{fig:a250mu600tb15}
in which $\mu=600$\,GeV.
This is because the sum of the cross sections of the processes $\n\n\rightarrow$
two bosons becomes
too small to obtain 
sufficiently small  relic density of the neutralino,
because 
the Higgsino components of the lightest neutralino become 
smaller. 
In the CMSSM, in addition to the difficulty in realizing small $\mu$,
the modes $\n\n\rightarrow$ two bosons except for $\n\n\rightarrow Zh$ does not open in  
small $\mhf$ region because of the heavier Higgs sector. 
In the black region,
the main mode is $\n\n\rightarrow bb$.
We comment on the $\mzr$ dependence of the relic density. 
When the sfermion mass becomes smaller, 
$\n\n\rightarrow$ two leptons process becomes larger because 
the cross section of the process via slepton t-channel exchange increases.
Therefore,
the smaller $\mzr$ leads to the larger annihilation cross section
and the smaller energy density of the lightest neutralino in the region.

\begin{figure}[!ht]
\begin{tabular}{cc}
\begin{minipage}{.5\textwidth}
\vspace{2mm}
\includegraphics[keepaspectratio,width=.9\textwidth]{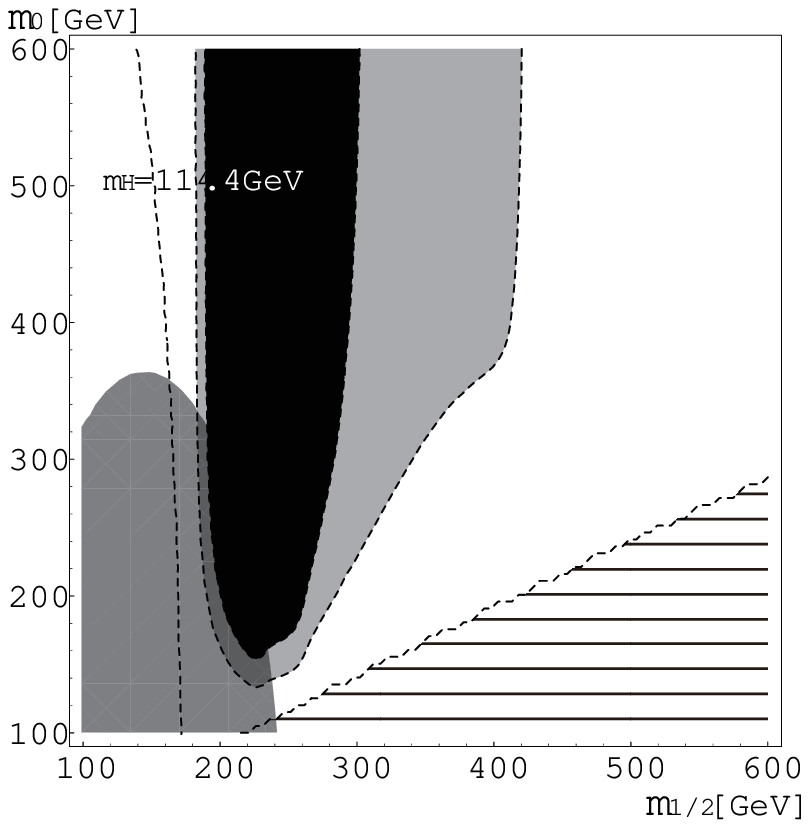}
\caption{$A_0=250$\,GeV, $\mu=325$\,GeV}
\label{fig:a250mu325tb15}
\end{minipage}&
\begin{minipage}{.5\textwidth}
\vspace{1mm}
\includegraphics[keepaspectratio,width=.92\textwidth]{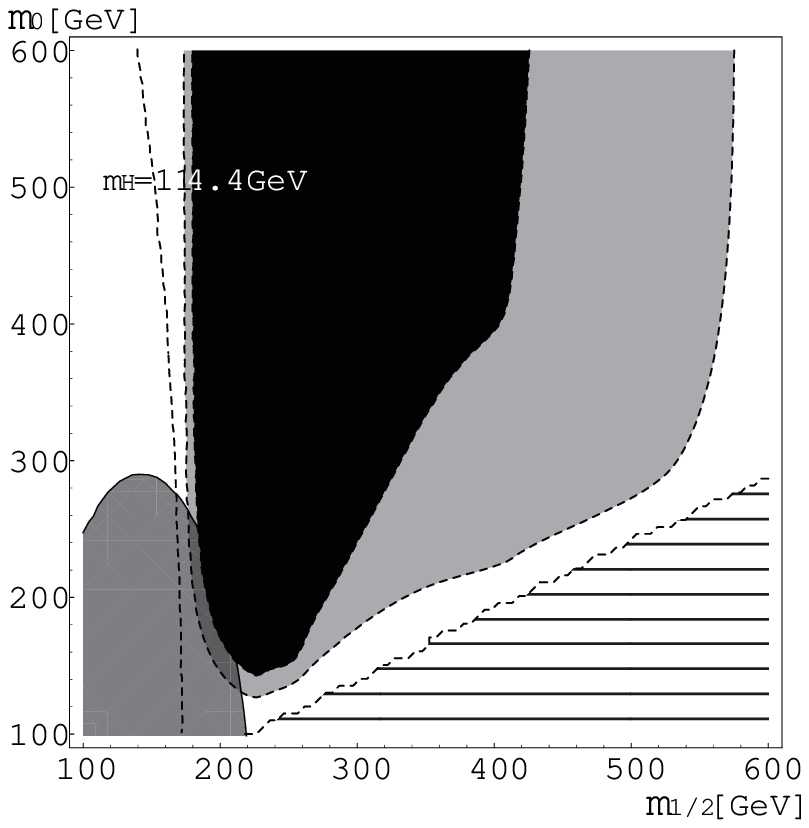}
\vspace{-2mm}
\caption{$A_0=250$\,GeV, $\mu=350$\,GeV}
\label{fig:a250mu350tb15}
\end{minipage}
\\ \multicolumn{2}{c}{\begin{minipage}{\textwidth}
\vspace{5mm}
\vbox{The ($\mhf$, $\mzr$) planes for $\mu=325$\,GeV (left) and $\mu=350$\,GeV (right) with $\Azr=250$\,GeV, $\tan{\beta}=15$ and $\ma=96$\,GeV. The usage of each color and line is the same as the previous figures.}
\end{minipage}}
\end{tabular}
\end{figure}

The graph of the relic density depends on the $\mu$ parameter strongly as we commented.
The distance between two allowed band region becomes wider 
in $\mu=300$\,GeV than in $\mu=275$\,GeV.
When $\mu$ is larger, the left preferred band moves to the left
because the lightest neutralino becomes heavier.
The right preferred band moves to the right,
because the coannihilation cross sections to two bosons becomes smaller for the larger $\mu$ parameter.
In this scenario,
the relic density of the neutralino does not depend on the gaugino mass so much 
when the gaugino mass is larger than $300$\,GeV in the parameter region we scanned.
Usually when the gaugino mass increases,
the total cross section decreases.
However,
in our scenario,
the Higgsino components of the neutralino increase,
and thus the two effects can almost compensate each other.
In Fig.\ref{fig:a250mu325tb15} and Fig.\ref{fig:a250mu350tb15} we can explicitly see the mild change of the relic density as the fairly broad prefferrered region 
when $\mu=325$ and $350$\,GeV.

The relic density does not change so much if we enlarge $\Azr$.
On the other hand, 
the cross section of the $b\rightarrow s\gamma$ process depends on $\Azr$ as
in the Fig.\ref{fig:a0mu275tb15}-Fig.\ref{fig:a250mu300tb15}.
There are reasonable regions which are consistent with the observed
relic density of dark matter and experimental constraints when $\Azr=250$\,GeV.
When $\Azr=0$\,GeV,
there are no or absolutely thin preferred region.
However, we do not take this allowed region for the $b\rightarrow s\gamma$ constraint seriously.
This is because the allowed region can be changed if there is other contributions to $b\to s\gamma$ as sizable gluino contribution.
Moreover,
the experimental value of the $Br(b\rightarrow s\gamma)$ is 
larger than the SM prediction.
In order to increase the $Br(b\rightarrow s\gamma)$,
larger SUSY breaking scale is required because the chargino contribution decreases the branching ratio.
This requirement makes the allowed region move to upper right. 

\section{Conclusions}
\label{sec:summary}
We have studied the thermal relic density of the neutralino
in the MSSM with the light Higgs sector and reasonable SUSY 
breaking parameters.
Actually, we took all the dimensionful parameters as order the
weak scale. 

The neutralino relic density with the light Higgs bosons
is totally different from the well-known result in the CMSSM.
In the natural light Higgs scenario,
the neutralinos to two bosons processes,
such as $\n\n\to hA$ and 
$\n\n\to HA$
open even when the gaugino mass is small.
Furthermore,
they dominate the total cross section
because the $\mu$ parameter is so small that  
the Higgsino components of the lightest neutralino
become comparatively large.
For those reasons,
the relic density become smaller than the observed value.
In contrast,
it is mostly larger than the observed value
in the CMSSM
because all Higgs bosons except for the SM-like Higgs are heavy.
The region with small relic density cannot be excluded
because it is not inconsistent if there is other dark matter 
sources.
Furthermore,
there are cosmologically preferred regions in this scenario,
which are not excluded by the experiments.
Thus the MSSM with the natural light Higgs sector is a good model 
not only for naturalness 
but also for cosmology.

The cosmologically preferred regions which we studied so far 
can be tested  by future direct searches for the weakly interacting massive particles (WIMP) because the small $\mu$ parameter makes the spin-independent interaction between Higgs and the lightest neutralinos large \cite{Ellis:2005mb,Asano},
unless recent direct searches for the WIMP
such as CDMSI\!I \cite{Akerib:2005kh} and XENON10 \cite{Angle:2007uj}
have excluded the regions.
In the region where the gaugino mass is roughly larger than $300$\,GeV,
even if the relic density is smaller than the observed relic density of
dark matter, it is not so small. (It is larger than $20\%$ of the observed value
in Fig.\ref{fig:a0mu275tb15}-Fig.\ref{fig:a250mu350tb15}.)
Therefore, even if the main component for dark matter around the galaxies
has only super weak interaction and cannot be found in the direct searches of the WIMP,
the searches can find the signal for the neutralino which is subdominant component,
though the concrete prediction becomes difficult.
We think it an interesting future subject
to study the direct detection of the WIMP in our scenario.

\section*{Acknowledgments}
S.-G.K, N.~M.,
K.~I.~N.,
and K.~S.\ are supported in part by Grants-in-Aid for Scientific 
Research from the Ministry of Education, Culture, Sports, Science, 
and Technology of Japan.
The work of T.~Y.\ was
supported by 21st Century COE Program of Nagoya University provided by JSPS.

\bibliography{basename of .bib file}

\end{document}